\documentclass[onecolumn,showpacs,preprintnumbers,amsmath,amssymb]{revtex4}
\usepackage{graphicx}
\usepackage{dcolumn}
\usepackage{bm}
\begin{document}
\preprint{Phys. Rev. {\bf D} 73, 053006 (2006)}
%%%%%%%%%%%%%%%%%%%%%%%%
\newcommand{\hs}{\hspace*{0.5cm}}
\newcommand{\vs}{\vspace*{0.5cm}}
\newcommand{\be}{\begin{equation}}
\newcommand{\ee}{\end{equation}}
\newcommand{\bea}{\begin{eqnarray}}
\newcommand{\eea}{\end{eqnarray}}
\newcommand{\ben}{\begin{enumerate}}
\newcommand{\een}{\end{enumerate}}
\newcommand{\bde}{\begin{widetext}}
\newcommand{\ede}{\end{widetext}}
\newcommand{\nn}{\nonumber}
\newcommand{\crn}{\nonumber \\}
\newcommand{\non}{\nonumber}
\newcommand{\noi}{\noindent}
\newcommand{\al}{\alpha}
\newcommand{\la}{\lambda}
\newcommand{\bet}{\beta}
\newcommand{\ga}{\gamma}
\newcommand{\va}{\varphi}
\newcommand{\om}{\omega}
\newcommand{\pa}{\partial}
\newcommand{\fr}{\frac}
\newcommand{\bc}{\begin{center}}
\newcommand{\ec}{\end{center}}
\newcommand{\Ga}{\Gamma}
\newcommand{\de}{\delta}
\newcommand{\De}{\Delta}
\newcommand{\ep}{\epsilon}
\newcommand{\varep}{\varepsilon}
\newcommand{\ka}{\kappa}
\newcommand{\La}{\Lambda}
\newcommand{\si}{\sigma}
\newcommand{\Si}{\Sigma}
\newcommand{\ta}{\tau}
\newcommand{\up}{\upsilon}
\newcommand{\Up}{\Upsilon}
\newcommand{\ze}{\zeta}
\newcommand{\ps}{\psi}
\newcommand{\Ps}{\Psi}
\newcommand{\ph}{\phi}
\newcommand{\vph}{\varphi}
\newcommand{\Ph}{\Phi}
\newcommand{\Om}{\Omega}
\def\lappeq{\mathrel{\rlap{\raise.5ex\hbox{$<$}}
{\lower.5ex\hbox{$\sim$}}}}
%%%%%%%%%%%%%%%%%%%%%%%%

\title{Interesting radiative patterns of neutrino mass\\  in
an
 $\mbox{SU}(3)_C\otimes \mbox{SU}(3)_L \otimes \mbox{U}(1)_X$
 model\\  with right-handed neutrinos}

\author{Darwin Chang}
%\email{pvdong@iop.vast.ac.vn}
\affiliation{ NCTS and Physics Department, National Tsing-Hua
University,
 Hsinchu 30043, Taiwan, ROC}
\author{Hoang  Ngoc Long}
\email{hnlong@iop.vast.ac.vn} \affiliation{ Physics Division,
NCTS, National Tsing-Hua University,
 Hsinchu 30043, Taiwan, ROC\\
 and\\
 Institute of Physics, VAST, P. O. Box 429, Bo Ho, Hanoi
10000, Vietnam }

\date{\today}

\begin{abstract}

We investigate a simple model of neutrino mass based on
$\mbox{SU}(3)_C\otimes \mbox{SU}(3)_L \otimes \mbox{U}(1)_X$ gauge
unification. The Yukawa coupling of the model has automatic
lepton-number symmetry which is broken only by the self-couplings
of the Higgs boson. At tree level neutrino spectrum contains three
Dirac fermions, one massless and two degenerate in mass. At the
two loop-level, neutrinos obtain Majorana masses and correct the
tree-level result which naturally gives rise to an inverted
hierarchy mass pattern and interesting mixing which can fit the
current data with minor fine-tuning.  In another scenarios, one
can pick the scales such that the loop-induced Majorana mass
matrix is bigger than the Dirac one and thus reproduces the usual
seesaw mechanism.

\end{abstract}


\maketitle

\section{Introduction}

The recent experimental results of SuperKamiokande
Collaboration~\cite{superK},  KamLAND~\cite{kam} and
SNO~\cite{sno} confirm that neutrinos have {\it tiny} masses and
oscillates.  This implies that the Standard Model (SM) of $
\mbox{SU}(2)_L \otimes \mbox{U}(1)_Y$ theory must be extended.

The solar and atmospheric neutrino oscillations are now firmly
established~\cite{nos}. The  $\De m^2$ values and mixing angles
are known with fair accuracy~\cite{gos,fogli}
 \bea  \De
m^2_{atm}&=& 2.4(1^{+0.21}_{-0.26})\times
10^{-3}~\textrm{eV}^2,\crn
       \De m^2_{sol}& = & 7.92(1 \pm 0.09)\times
       ~10^{-5}~\textrm{eV}^2,\crn
 \sin^2 {\theta_{23}} & = &   0.44 (1^{+0.41}_{-0.22})
 , \  \sin^2 {\theta_{12}} = 0.314(1^{+0.18}_{-0.15}),\crn
 \sin^2 {\theta_{13}} &=& 0.9^{+2.3}_{-0.9} \times 10^{-2}.
\eea
 The tritium experiments~\cite{tri} provide an upper bound on the
absolute value of neutrino mass \be m_i \leq 2.2 \ \textrm{eV}\ee
A more strict bound
\[ m_i \leq 0.6\ \textrm{eV}\]
was found from the analysis of the latest cosmological
data~\cite{cos}.

Since the data only provide the information about difference in
$m_{\nu}^2$, the neutrino mass pattern can either be almost
degenerate, or hierarchical. Among the hierarchical possibilities,
there are two types: normal hierarchical or inverted hierarchical.
In the literature, most of the models explore normal hierarchical
case.

In this paper we will explore a model which naturally gives rise
to three pseudo-Dirac neutrinos with inverted hierarchical mass
pattern.

Among the possible extensions of the SM, a curious choice are the
3-3-1 models which are based on the simplest non-Abelian extension
of the SM group, namely, the $\mbox{SU}(3)_C\otimes \mbox{SU}(3)_L
\otimes \mbox{U}(1)_X$~\cite{ppf,flt}. The reason why these models
are appealing has been exposed in many recent
publications~\cite{recent}. The model requires that the number of
fermion families be a multiple of the quark color in order to
cancel anomalies, which suggests an interesting connection between
the number of flavors and the strong color group.
 If one further  uses the
condition of QCD asymptotic freedom, which is valid only if the
number of families of quarks is to be less than five, it follows
that $N$ is equal to 3. In addition, the third quark generation
has to be different from the first two, so this leads to the
possible explanation of why top quark is uncharacteristically
heavy.

There are two main versions of the 3-3-1 models as far as lepton
sector is concerned. In the minimal version, the charge
conjugation of the right-handed charged lepton for each generation
is combined with the usual $SU(2)_L$ doublet left-handed leptons
components to form an $SU(3)$ triplet $(\nu, l, l^c)_L$.  No extra
leptons are needed and there we shall call such models minimal
3-3-1 models. There is no right-handed neutrino in its minimal
version. Another version adds a left-hand antineutrino to each
usual $SU(2)_L$ doublet left-handed lepton to form a triplet, {\it
i. e.}, $(\nu, l, \nu^c)_L$~\cite{flt}.  These left-handed
anti-neutrinos serve the role of the charge conjugation of the
usual right-handed neutrinos which are required in the usual
seesaw mechanism. We therefore call such models right-handed
neutrino models (RHNM). It is this type of model that we shall
explore in this manuscript. Its main feature is that it requires
only a more economic Higgs sector for breaking the gauge symmetry
and generating the fermion mass. Among the new gauge bosons of
this model, the non-self-conjugated neutral boson $X^0$ can have
promising signature in accelerator experiments and it can also be
the source of neutrino oscillations~\cite{til}.

 The explanation of the smallness of the neutrino masses and
the profile of their mixing as required by recent experiments have
been a great puzzle in particle physics. In the past several years
a great amount of papers have been devoted to its solution (on the
neutrino mass in the minimal 3-3-1 model, see
Refs.~\cite{om,yasue,nmass}).

The most popular mechanism is of course the seesaw model with a
few very heavy right-handed, $SU(2)_L$ singlet, neutrinos. This
type of model requires a new very high scale of $10^{12} $  GeV or
higher. An alternative mechanism for generating small neutrino
masses, which may not requires such high scale, is to do it only
as one or multiloop radiative corrections. In the framework of
$SU(2) \otimes U(1)$ model, a famous example is the so-called Zee
Model and its generalizations~\cite{cp,cz}. In the framework of
the minimal 3 - 3 - 1 model, this mechanism has been considered
in~\cite{yasue} based on the Zee type mechanism i. e. by
introducing a scalar singlet.

 In this paper we  shall explore the alternative RHNM in its minimal
form. It is shown, with minimal Higgs sector, that the Yukawa
sector has automatic lepton-number conservation which is broken in
the Higgs sector. At tree level the neutrino spectrum contains
three Dirac fermions, one massless and two degenerate in mass. At
the two-loop level, very much like one of the Zee Models, with the
help of lepton-number violating Higgs couplings, neutrinos obtain
Majorana masses and correct the tree-level result. Since the
Majorana masses involve a new physics ($SU(3)_L$ breaking) scale,
depending on the size of the scale there are two scenarios
possible.  In the first one, the $SU(3)_L$ breaking scale is
chosen to be very high and as a result the right-handed Majorana
mass matrix is still very large, even though it is two
loop-induced, compared with the Dirac mass.  In this case, the
usual seesaw mechanism still applies. In another scenario, the
$SU(3)_L$ breaking scale is chosen to be not much higher than the
weak scale; in that case the following interesting pattern arises.
This naturally gives rise to an interesting inverted hierarchy
mass pattern and interesting mixing which can fit the current data
with some fine-tuning (to make the tree-level Dirac mass of order
$\Delta m^{\nu}_{atm}$). This radiative correction naturally
occurs without introducing extra scalar singlet that was  needed
in the minimal model~\cite{yasue}. This scenario gives rise to a
pseudo-Dirac neutrino mass pattern. There are many discussions of
pseudo-Dirac neutrino mass pattern in the
literature~\cite{pseudo,kl,pak}, however our scenarios is
different from all of them as we will discuss. Throughout the
paper we shall try to keep each sector minimal and see what kind
of neutrino pattern is produced in general in the context of RHNM.
We shall not implement by hand any extra texture in order to
generate special pattern that can fit data.

This paper is organized as follows. In Sec. \ref{secmodel} we
review the 3-3-1 model with right-handed neutrinos and introduce
the Higgs content and the Yukawa couplings. The conserved charges
$\mathcal{L}$ and $\mathcal{B}$ are introduced and lepton-number
violating couplings in the scalar sector are discussed and the
general mass matrix is presented in Sec. \ref{radia}, while in
Sec. \ref{twoloopcal} we derive the mass matrix through two-loop
corrections. The main neutrino  phenomena are   are presented in
Sec. \ref{pheno}. Finally, the last section is devoted to our
conclusions.

\section{The  3-3-1 model with right-handed neutrinos}
\label{secmodel}
 To frame the context, it is appropriate to
recall briefly some relevant features of the 3 - 3 - 1 model with
right-handed neutrinos~\cite{flt}. In this model, the leptons are
in triplets, in which the third member is a right-handed neutrino:
\be f^{a}_L = \left(
               \nu^a_L, l^a_L, N^a_L
\right)^T \sim (1, 3, -1/3), l^a_R\sim (1, 1, -1),
  \label{l2}
\ee where $a = 1, 2, 3$ is a  family index. Here the right-handed
neutrino is denoted by $ N_L  \equiv   (\nu_R)^C$. Note  the fact
that there are three generations of leptons is a peculiar
consequence of anomaly cancellation as discussed in the
introduction.  This is an interesting plus to this type of models.
The first two generations of quarks are in antitriplets while the
third one is in a triplet and each charged left-handed fermion
field has its right-handed counterpart transforming as a singlet
of the SU(3)$_L$ group
 \be Q_{\al L} = \left(
                d_{\al L},-u_{\al L}, D_{\al L},\\
                 \right)^T \sim (3, \bar{3}, 0), \  \al=1,2
\label{q}, \ee \be
 Q_{3L} = \left(
                 u_{3L}, d_{3L}, T_{L}
                 \right)^T \sim (3, 3, 1/3),
 T_{R} \sim (3, 1, 2/3),\ee
\bea D_{\al R}&\sim & (3, 1, -1/3),  u_{a R}\sim (3, 1, 2/3),\crn
d_{a R}& \sim& (3, 1, -1/3),
 \ a=1,2,3.
\eea
 Note that the five quarks $d_{a R}$ and $D_{\al R}$ have the
same quantum number and so are the four quarks $u_{a R}$ and
$T_{R}$. Their identity are defined only by the convention of the
Yukawa couplings that we adopt as will be explained later. Also
note that the third generation has different gauge content
compared with the first two generations which is required by the
anomaly cancellation. The electric charge operator is given in the
form \be Q=\fr{\la_3}{2} - \fr{\la_8}{2\sqrt{3}} + X, \ee where
$X$ is the $U(1)$ gauge charge, $\la_i$ are the $SU(3)_L$ gauge
charge. The non-self-conjugated gauge bosons are defined as \bea
\sqrt{2}\ W^+_\mu &=& W^1_\mu - iW^2_\mu ,
\sqrt{2}\ Y^-_\mu = W^6_\mu - iW^7_\mu ,\nn\\
\sqrt{2}\ X_\mu^o &=& W^4_\mu - iW^5_\mu, \eea where $W^i$ are the
gauge boson associated $\la_i$. The {\it physical} neutral,
self-conjugated gauge bosons associated with generator $\lambda_3$
and $\lambda_8$ and  $X$, besides the photon, are again related to
$Z, Z'$ through the mixing angle $\phi$.

 The gauge symmetry breaking and fermion mass generation can be
achieved with just three $\mbox{SU}(3)_L$ triplets \bea \rho & =&
\left(
                \rho^+_1,\  \rho^0_2,\  \rho^{+}_3
                  \right)^T \sim (1, 3, 2/3),\label{hig}\nn\\
\eta & =& \left(
                \eta^0_1,\  \eta^-_2,\  \eta^{0}_3
                 \right)^T \sim (1, 3, -1/3),\\
\chi & = &\left(
                \chi^0_1,\  \chi^-_2,\  \chi^{0}_3
                 \right)^T \sim (1, 3, -1/3).\nn
\eea Note that the scalars $\eta$ and $\chi$ have the same quantum
numbers.  By convention, we define $\chi$ to be the one with
nonzero $\langle \chi_3^0 \rangle$ and breaks $SU(3)_L$, therefore
$\langle \eta_3^0 \rangle = 0$ by convention. The necessary VEVs
are
 \bea \langle\rho \rangle & = & (0, u/ \sqrt{2},\  0)^T,\
\langle\eta \rangle  = ( v/ \sqrt{2},\  0,\ 0)^T, \crn \langle\chi
\rangle & = & (0,\ 0,\omega/\sqrt{2})^T. \label{vev} \eea
 In general, $\langle
\chi_1^0 \rangle$ can also be nonzero, however, its effect is
small and we shall ignore it in the following. Note that the
identity of $\chi$ is defined by convention to be the one will be
responsible for $SU(3)_L$ breaking while $\rho$ and $\eta$ are
responsible for $SU(2)_L$ breaking. The reason why two triplets
are needed for $SU(2)_L$ breaking is because the three generations
have different gauge charge, and one triplet is not enough to give
fermion mass to all three generations (see below).

The most general Yukawa Lagrangian as follows: \bde
\begin{eqnarray}
{\cal L}_{Y}^{\chi}&=&h_1\bar{Q}_{3L}T_{R}\chi +
 h_{2\al \bet }\bar{Q}_{\al L}D_{\bet R}\chi^{*} + \mbox{h.c.}\crn
 &=&h_1(\bar{u}_{3L}\chi_1^0+\bar{d}_{3L}\chi_2^-
+\bar{T}_L\chi_3^{0})T_R +h_{2\al \bet}(\bar{d}_{\al
L}\chi_1^{0*}-\bar{u}_{\al L}\chi_2^+ + \bar{D}_{\al
L}\chi_3^{0*})D_{\bet R} + \mbox{h.c.}\crn
{\cal L}_{Y}^{\eta}&=&h_{3a}\bar{Q}_{3L}u_{aR}\eta+ h_{4\al
a}\bar{Q}_{\al L}d_{aR}\eta^{*}+\mbox{h.c.}\crn
&=&h_{3a}(\bar{u}_{3L}\eta_1^0+\bar{d}_{3L}\eta_2^- +
\bar{T}_L\eta_3^{0}) u_{aR}+h_{4\al a}(\bar{d}_{\al
L}\eta_1^{0*}-\bar{u}_{\al L}\eta_2^+
+\bar{D}_{\al L}\eta_3^{0*})d_{aR}+\mbox{h.c.}\nonumber\\
{\cal L}_{Y}^{\rho} &=& h_{5a}\bar{Q}_{3L}d_{aR}\rho +
 h_{6\al a}\bar{Q}_{\al L}u_{aR}\rho^{*}+G_{ab}\bar{f}^a_L l^b_R\rho+
F_{ab}\varep_{ijk}(\bar{f}_{L})^{a i}(f^C_L)^{b j}
(\rho^{*})^k+\mbox{h.c.}\crn &=&h_{5a}(\bar{u}_{3L}\rho_1^+ +
\bar{d}_{3L}\rho_2^0 + \bar{T}_L\rho_3^{+})d_{aR} + h_{6\al
a}(\bar{d}_{\al L}\rho_1^- -\bar{u}_{\al
L}\rho_2^{0*}+\bar{D}_{\al L}\rho_3^{-})u_{aR}\crn & &+
G_{ab}[\overline{\nu}^a_L\rho_1^+ +
\bar{l}^a_L\rho_2^0+\overline{N}_L^a \rho_3^{+}]l^b_R+\crn
&&+F_{ab}\{\overline{\nu}^a_L[(l^C_L)^b\rho_3^{-} -(N_L^C)^b
\rho_2^0] +\bar{l}^a_L[(N^C_L)^b\rho_1^{-} -
(\nu^C_L)^b\rho_3^-]\crn &&+ {\overline N}_L^a[(\nu^C_L)^b\rho_2^0
- (l_L^C)^b\rho_1^-] \}   + \mbox{h.c.}. \label{yukawa1} \eea \ede
Note that, by convention, $T_{R}$ is defined to be the one that
couples to $\bar{Q}_{3L} \chi$ among the four quarks with the same
quantum numbers, similarly, $D_{\bet R}, (\beta=1,2) $, are
defined to be the two quarks that couple to $\bar{Q}_{\al L} \chi$
among the five quark with the same quantum numbers.  Note that
$G_{ab}$ gives rise to charged lepton Dirac masses,  while
$F_{ab}$, which is antisymmetric, gives rise to the Dirac masses
for neutral leptons.

 The leptons have the Yukawa couplings only with the $\rho$ Higgs
boson.   One can find a naive lepton number $L_N$
 is violated only through the $F_{ab}$
coefficients while the rest of the whole Lagrangian, including the
Yukawa couplings $G_{ab}$, is  the lepton-number  conserving. Only
the leptons carry the $L_N$ charge: $L_N(l^a_R)=1, L_N(f^a_L)=1$.
Since phenomenologically, one requires $F_{ab}$ to be much smaller
than $G_{ab}$, it can be done in our context only by fine-tuning.
The $L_N$ allows us to claim that this fine-tuning is technically
natural (in t'Hooft sense).  We call $L_N$``naive" because it
defines the lepton number $\nu_R^C = N_L$ to be different from
$\nu_L$.
 Later, we will introduce another lepton number, $L$,
 with $L(\nu_L) = L(\nu_R)$ like the conventional lepton number,
 which will play an important role in our discussion on neutrino masses.
The  lepton Yukawa couplings needed in this work are presented in
Fig. \ref{figdl1}. \vs

\begin{figure*}[htbp]
\begin{center}
\includegraphics[width=12cm,height=8cm]{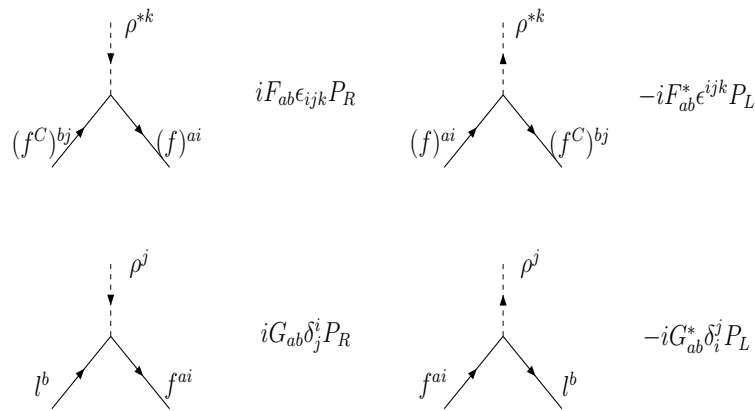}
\caption{\label{figdl1}{The necessary Yukawa couplings }}
\end{center}
\end{figure*}

 The VEV $\langle\chi \rangle$ breaks $SU(3)_L \times U(1)_X$
down to $SU(2)_L \times U(1)_Y$ and gives masses of the exotic
quarks as well as non-SM gauge bosons $X,Y$ and $Z'$. The VEV
$\langle\eta \rangle$ gives mass for $u_3, d_{\al}$ quarks, while
$\langle\rho \rangle$ gives mass for  $u_\al, d_{3}$ and {\it all}
ordinary leptons. The SM gauge bosons gain mass both from VEVs of
$\eta$ and $\rho$.

  After symmetry breaking the gauge bosons gain  masses \bea
m^2_W &= &\frac{1}{4}g^2(u^2+v^2),\
M^2_Y=\frac{1}{4}g^2(v^2+\omega^2),\crn M^2_X &=
&\frac{1}{4}g^2(u^2+\omega^2). \label{rhb} \eea $W^4$ and $W^5$
accidentally have the same mass. Eq.(\ref{rhb}) implies $ v_W^2 =
u^2 + v^2 =246^2\ \  \mbox{GeV}^2$.  In order to be consistent
with the low energy phenomenology we have to assume that $\langle
\chi \rangle \gg\ \langle \rho \rangle,\ \langle \eta \rangle$,
such that $m_W \ll M_X, M_Y$. The symmetry-breaking hierarchy
gives us splitting on the bilepton masses~\cite{til}  $ | M_X^2 -
M_Y^2 | \leq m_W^2$.  Since $m_W \ll M_X, M_Y$,  we can take $M_X
\approx M_Y$. The ``wrong'' muon  decay limit
\[
  R = \frac{\Gamma (\mu^- \rightarrow e^- \nu_e \bar{\nu}_\mu)}
{\Gamma (\mu^- \rightarrow e^- \bar{\nu}_e \nu_\mu)} \sim
\left(\frac{m_W}{M_Y}\right)^4 < 1.2\%, \ 95 \% \   \textrm{CL}
\]
gives  $ M_{Y^-} \ge 230 $ GeV. From consideration of muon decay
parameters, one has got the mass bound of the singly-charged
bilepton of $440 $ GeV ~\cite{tj98}. With this mass scale,
$\langle\chi \rangle \sim  800 $ GeV.

\section{Radiative corrections to neutrino mass}
\label{radia}

 The Yukawa couplings of Eq.(\ref{yukawa1}) possess extra global
symmetries which are not broken by VEVs $u, v, \omega$. From the
Yukawa couplings, one can find the following lepton symmetry $L$
as in Table \ref{lnumber} (only the fields with nonzero $L$ is
listed, all other fields have vanishing L).
\begin{table*}
\caption{
    Nonzero lepton number $L$ of fields in the 3-3-1 model with RH neutrinos.}
\begin{center}
\begin{tabular}{c|ccc|cccc|cccc|}
    \hline
        Fields
&$N_L$&$l_L$&$l_R$
& $\rho^+_3$&$\eta^0_3$&$\chi^0_1$&$\chi^-_2$
&$D_{\alpha L}$& $D_{\beta L}$&$T_L$&$T_R$\\
    \hline
        $L$ & $-1$ & $1$ & $1$ & $-2$&$-2$&$2$&$2$&$2$&$2$&$-2$&$-2$ \\
    \hline
\end{tabular}
\label{lnumber}
\end{center}
\end{table*}

It is interesting that the exotic quarks also carry the lepton
number. However, this $L$ obviously does not commute with gauge
symmetry.  One can construct a new conserved charge $\cal L$
through $L$ by making the linear combination $L= x\la_3 + y\la_8 +
zX + {\cal L} I$ where $\la_3$ and $\la_8$ are $SU(3)_L$
generators.
 One finds the following solution (see also \cite{tj})
$x=0, y= \fr{2}{\sqrt{3}}, z=0$, and \be L = \fr{2}{\sqrt{3}}\la_8
+ {\cal L} I \label{lepn} \ee as in Table~\ref{bcharge}. Another
useful conserved charge $\cal B$ is usual baryon number $B ={\cal
B} I$.
\begin{table*}
%\begin{table}[h]
\caption{
     ${\cal B}$ and ${\cal L}$ charges for multiplets in
the 3-3-1 model with RH neutrinos.}
\begin{center}
\begin{tabular}{l|ccccccccccc}
Multiplet & $\chi$ & $\eta$ & $\rho$ &  $Q_{3L}$ & $Q_{\al L}$ &
$u_{aR}$&$d_{aR}$ &$T_R$ & $D_{\al R}$ & $f_{aL}$ & $l_{aR}$  \\
\hline $\cal B$ charge &$0$ & $ 0  $ &
 $ 0  $ &  $\fr 1 3  $ & $\fr 1 3  $& $\fr 1 3  $ &
 $\fr 1 3  $ &  $\fr 1 3  $&  $\fr 1 3  $&
 $0  $& $0   $\\
\hline $\cal L$ charge &$\fr 4 3$ & $-\fr 2 3  $ &  $-\fr 2 3  $ &
   $-\fr 2 3  $ & $\fr 2 3  $& 0 & 0 & $-2$& $2$&
 $\fr 1 3  $& $ 1   $\\
\end{tabular}
\label{bcharge}
\end{center}
%\end{table}
\end{table*}

Note that even though $\eta$ and $\chi$ triplets have the same
quantum number, they are distinguished  already by our convention
of Yukawa couplings and VEV's and, as a result, their lepton
number assignments are {\it quite different}: $\eta_1^0$ and
$\chi^0_3$ do not have lepton number $L=0$, while $\eta_3^0$ and
$\chi^0_1$ are  bilepton $L=2$.

 The lepton number ${\cal L}$ is, however, broken
 in the Higgs potential in
general.  The most general potential can then be written as
the sum of the ${\cal L}$ conserving $V_{LNC}$ and
${\cal L}$ violating $V_{LNV}$  (see also~\cite{rrf}):

\be V(\eta,\rho,\chi) = V_{LNC}(\eta,\rho,\chi) +
V_{LNV}(\eta,\rho,\chi),\ee where $V_{LNC}(\eta,\rho,\chi)$ is
\bde \bea V_{LNC}(\eta,\rho,\chi)&=&\mu^2_1 \eta^+ \eta +
 \mu^2_2 \rho^+ \rho +  \mu^2_3 \chi^+ \chi +
\lambda_1 (\eta^+ \eta)^2 + \lambda_2 (\rho^+ \rho)^2 \crn && +
\lambda_3 (\chi^+ \chi)^2  + (\eta^+ \eta) [ \lambda_4 (\rho^+
\rho) + \lambda_5 (\chi^+ \chi)] + \lambda_6 (\rho^+ \rho)(\chi^+
\chi) + \lambda_7 (\rho^+ \eta)(\eta^+ \rho) \crn && + \lambda_8
(\chi^+ \eta)(\eta^+ \chi) + \lambda_9 (\rho^+ \chi)(\chi^+ \rho)
+ [ \mu_5 \ep^{ijk}\eta_i\rho_j\chi_k + h.c ] \label{pot1c}, \eea
and \bea V_{LNV}(\eta,\rho,\chi)&=&  \overline{\mu_4}^2 (\chi^+
\eta +
 \eta^+ \chi) +
 (\eta^+ \chi)[{\overline \la}_{11} (\rho^+\rho) +
\overline{\la}_{12}(\eta^+\eta) +\overline{\la}_{13}(\chi^+\chi)]
+ h.c\crn
 &&+\overline{\la}_{10} (\chi^+ \eta + \eta^+ \chi)^2
+ [\overline{\la}_{14} \eta^+\rho\rho^+\chi
 + h.c ],
\label{pot1v} \eea \ede
 where overbars have been used to denote
lepton-number violating couplings. The Higgs boson couplings
necessary in this work are depicted in Fig. \ref{figdl2}. \vs

\begin{figure*}[htbp]
\begin{center}
\includegraphics[width=12cm,height=5cm]{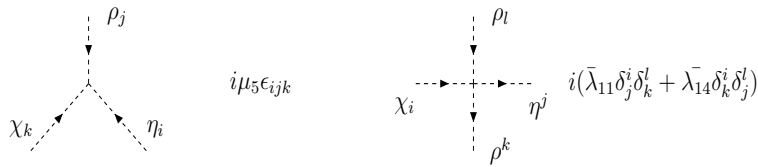}
\caption{\label{figdl2}{ The necessary Higgs boson couplings  }}
\end{center}
\end{figure*}

At tree level the neutrinos get
Dirac masses from the Yukawa coupling $F_{ab}\varep_{ijk}$
$(\bar{f}_{L})^{a i}$ $(f^C_L)^{b j} (\rho^{*})^k$.  One can
always assume that $G_{ab}$ is diagonal by convention and
 can always pick fermion phases so that the three coefficients
$F_{12}, F_{13}, F_{23}$ are {\it all real}. The resulting Dirac
mass matrix is traceless and antisymmetric, and therefore has the
mass pattern $0, - m_\nu, m_\nu$. This is clearly not realistic.
However, this pattern is severely changed by the quantum effect.
In the base of $(\nu_e, \nu_\mu, \nu_\tau, N_e, N_\mu, N_\tau)_L$,
the most general mass matrix can be written as \bde \bea M_{\nu N}
=\left(
\begin{array}{cccccc}
 &&&  0 &F_{12}& F_{13}
\\
 &M_{\nu }/ \langle \rho_2 \rangle  & &-F_{12}&0&F_{23}
\\
  & & &-F_{13}&-F_{23}&0
  \\
 0 &-F_{12}&-F_{13}& & &
\\

 F_{12}&0&-F_{23}& &M_{ N}/ \langle \rho_2 \rangle   &
 \\
 F_{13}&F_{23}&0 & & &

\end{array}
\right) \langle \rho_2 \rangle,
 \label{matmn1}
 \eea
 \ede
where $M_{ \nu}$ and $M_{ N}$ can arise from quantum correction.
In particular, $M_{ \nu}$ can be due to the loop-induced operator
 \be
O_{\nu}(M_{\nu }) \sim  \fr{1}{M}(f_i
f_j)(\eta^{+})^i(\eta^{+})^j, \label{eop1}\ee
 which is lepton number violating
interaction, while $M_{ N}$ is due to
 \be O_N(M_{N }) \sim  \fr{1}{M}(f_i
f_j)(\chi^{+})^i(\chi^{+})^j.\label{eop2} \ee
The Dirac masses can also receive quantum
 correction from the lepton number
conserving operator \be O_d(M_{d}) \sim \fr{1}{M}(f_i
f_j)(\chi^{+})^i(\eta^{+})^j. \label{eop3}\ee In Ref.
\cite{dias}, the effective dimension-five operators (${\cal
O}_\nu$, ${\cal O}_N$) and (${\cal O}_d$) were used to obtain the
neutrino mass matrix. Choosing the free parameters in above
operators ($f, h$ and $g$) and taking $v_\eta = 10^2$ GeV, $v_\chi
= 10^3$ GeV and $\La = 10^{14}$  GeV, one have got neutrino masses
($m_1$, $m_2$, ... $m_6$) in the range $ 10^{-5} \div 1.7$ eV. The
authors argued that this set of parameters accommodates  the solar
and atmospheric oscillation data along with the  LSND experiment
altogether.

It is well known that the neutrinos can get a mass through
radiative mechanism~\cite{cp}. In the current model, neutrinos can
get mass through two-loop radiative corrections, which is
represented by the  Feynman diagrams depicted in Fig.\ref{figdl3}.

Since $ \langle \eta \rangle \ll \langle \chi \rangle$ it is
obvious that $M_\nu$ is small and negligible. The quantum
corrections to the Dirac mass terms are also clearly smaller and
negligible for the same reason [$\propto \langle \eta \rangle^2$,
see a notice after Eq.(\ref{p})]. However, radiative contributions
to $M_{N }$ can be very large and play a major role in determining
the neutrino mass pattern in this model. The size of $M_N$ depends
on the scale of $SU(3)_L$ breaking, $\langle \chi \rangle =
\omega/\sqrt{2}$, and the dominant scale in the loop $M$ in above
equations.  If $\omega/M$ is tuned to be very large, such as
$10^{12} $ GeV, then the $M_N$ can be much larger than the tree
level Dirac mass matrix and the usual seesaw mechanism will still
apply.  (In fact, in this case the loop corrections to the Dirac
mass may also have to be taken into account).  This scenario is
more standard and requires very large $\omega/M$ which is not
natural in our context.  In the following we will concentrate more
on the second, more natural, scenario in which $\omega/M$ is not
so large.  In that case, $M_N$ can be considered a small
correction to the dominant tree level Dirac mass.  A curiously
interesting neutrino mass pattern emerges. \vs

To finish this section, we mention that  in Ref. \cite{dias}, the
mass matrix  obtained from dimension-five effective operators in
(\ref{matmn1}) can give the possibility of explaining the LSND
data as well as solar and atmospheric neutrino oscillation.
 On the other hand, depending on the temperature at which light sterile
 neutrinos thermalize, they can play an important role in big-bang
 nucleosynthesis (BBN)~\cite{bbn} or other aspects of cosmological
 problems~\cite{cosm}. All these problems should be further studied
 but it is out of the scope of the present work.

\section{Two-loop corrections}
\label{twoloopcal}

Radiative correction stars only from the two-loop level. The most
important two-loop contributions (to $M_N$) turn out to happen at
the two-loop level.  They are shown in Fig.\ref{figdl3} \vs

\begin{figure*}[htbp]
\begin{center}
\includegraphics[width=13cm,height=9cm]{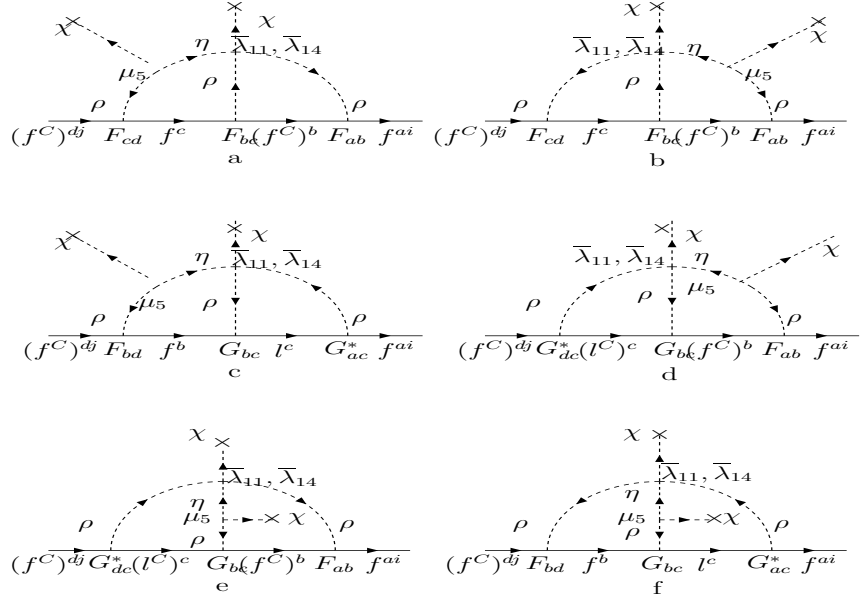}
\caption{\label{figdl3}{Two-loop contribution to neutrino mass
matrix }}
\end{center}
\end{figure*}

To calculate these diagrams, take \ref{figdl3} as an example,
using vertices in Fig.\ref{figdl1}, and Fig.\ref{figdl2}, the
contribution of diagram \ref{figdl3}a is given by \bde
 \bea
M_N^{F.3a}&=&\int \fr{d^4 q}{(2 \pi)^4} \fr{d^4 k}{(2 \pi)^4} i
F_{ab} \ep_{iki'} P_R \fr{i(-q\!\!/ + m_b)}{q^2 - m_b^2}(-i)
F_{bc}^* \ep^{klm} P_L \fr{i(k\!\!\!/ + m_c)}{k^2 - m_c^2}i F_{cd}
\ep_{ljt} P_R \crn & & \times i(\bar{\la}_{11} \de_n^3 \de_m^{i'}
+ \bar{\la}_{14} \de_m^3 \de_n^{i'}) \mu_5 \ep^{nt3}\fr{i}{(q^2 -
m^2_\rho)}\fr{i}{(k^2 - m^2_\eta)}\fr{i}{(k^2 -
m^2_\rho)}\fr{i}{[(k-q)^2 - m^2_\rho]}\crn &=& - 2 \bar{\la}_{14}
\de_i^3 \de_j^{3}\mu_5F_{ab}F_{bc}^*F_{cd}P_R \int \fr{d^4 q}{(2
\pi)^4} \fr{d^4 k}{(2 \pi)^4}\left[ \fr{(-q\!\!/)}{q^2 -
m_b^2}\fr{k\!\!\!/}{k^2 - m_c^2}\right.\crn && \times
\left.\fr{1}{(q^2 - m^2_\rho)}\fr{1}{(k^2 - m^2_\eta)}\fr{1}{(k^2
- m^2_\rho)}\fr{1}{[(k-q)^2 - m^2_\rho]}\right]P_R.
\label{dlt}\eea \ede
 From  (\ref{dlt}) we see that only the
neutrinos ($i=j=3$), and not the leptons, received Majorana
masses. Note that the integral in (\ref{dlt}) is {\it finite}.
 The contribution from both Figs. \ref{figdl3}(a) and \ref{figdl3}(b)
  (we call these F-type contributions) to the neutrino mass
matrix is given
 \bea M^{(F)}_{N(ai,dj)}& =& - 2 \de_{3i} \de_{3j} \mu_5 {\overline
\la}_{14}  \om^2 \sum_{b c} F_{ab} F_{bc}^+ F_{cd} A(b,c)\crn
&\equiv & a_{ij} \sum_{b c} F_{ab} F_{bc} F_{cd} A(b,c)
\label{cota}\eea where it has been denoted $a_{ij} \equiv - 2
\de_{3i} \de_{3j} \mu_5 {\overline \la}_{14} \om^2$ and \be A(b,c)
=  [ I_{1}(m^2_b, m^2_c) - I_{1}(m^2_c, m^2_b)]=-A(c,b) .\ee The
factor 2 in $a_{ij}$ is due to summation over the $SU(3)$ indexes.

 Here
 \bea I_{1}(m^2_b, m^2_c)& =& \int
\fr{d^4 k}{(2\pi)^4} \fr{d^4 q}{(2\pi)^4} \fr{k\!\!\!\!/
}{(k^2-m^2_c)} \fr{1}{(k^2-m^2_\rho)}\crn && \times \fr{q\!\!\!/
}{(q^2-m^2_b)} \fr{1}{(q^2-m^2_\rho)}\crn && \times
\fr{1}{[(k-q)^2-m^2_\rho]} \fr{1}{(k^2-m^2_\eta)}. \label{twoloop}
\eea
 It is easily to check out that
\bea M_{N}^{(F)} \simeq n_2\left( \begin{array}{ccc} 0 & F_{12}J_1
& F_{13}J_2 \\  F_{12}J_1 & 0 & F_{23}J_3 \\ F_{13}J_2 & F_{23}J_3
&0
\end{array} \right), \label{matn}\eea  where
 \bea n_2 & \equiv & 2 \mu_5 \bar{
\la}_{14} \om^2, \label{f}\\
 J_1 &=& F_{12}^2 A(1,2) + F_{13}^2 A(1,3)
- F_{23}^2 A(2,3),\nn \eea \bea
 J_2 &=& F_{12}^2 A(1,2) + F_{13}^2 A(1,3) +
F_{23}^2 A(2,3),\crn
 J_3 &=& -F_{12}^2 A(1,2) + F_{13}^2 A(1,3) +
F_{23}^2 A(2,3). \eea

 We can approximate the integral by
  \be A(b,c)
\approx  \left(\fr{1}{16 \pi^2}\right)^2 \fr{(m_b^2 -
m_c^2)}{{\overline M}^4}
 \label{abc}\ee
 where
${\overline M}$ is the dominant mass scale in the loop:
 ${\overline M} \approx m_\rho \approx m_\eta$.

The contribution from  Figs.\ref{figdl3}(c) and \ref{figdl3}(d)
(we call these G-type contributions) to the neutrino mass matrix
is given \bde
 \bea
M^{(G1)}_{ai,dj}& = &  a_{ij} \sum_{b, c} \left[F_{ab} G_{bc}
(G_{cd}^+) + F_{db} G_{bc} (G_{ca}^+)\right] \crn &  & \times \int
\fr{d^4 k}{(2\pi)^4} \fr{d^4 q}{(2\pi)^4}
\fr{k\!\!\!\!/}{(k^2-m^2_b)}
\fr{1}{(k^2-m^2_\rho)}\fr{1}{(k^2-m^2_\eta)}\fr{q\!\!\!/}{(q^2-m^2_c)}
\fr{1}{(q^2-m^2_\rho)} \fr{1}{[(k-q)^2-m^2_\rho]} \crn
 \eea
\ede
 Since $m_b, m_c \ll m_\rho, m_\eta$, the loop integral depends
on $m_b, m_c$ only weakly.  We see that the contribution is
approximately  proportional to $ (FGG^*)_{ad}$. Similarly, the
contribution from  Figs.\ref{figdl3}(e) and \ref{figdl3}(f) to the
neutrino mass matrix is given by \bde \bea M^{(G2)}_{ai,dj}& =  &-
a_{ij}\sum_{b, c} \left[F_{ab} G_{bc} (G_{cd}^+) + F_{db} G_{bc}
(G_{ca}^+)\right] \crn & &\times \int \fr{d^4 k}{(2\pi)^4} \fr{d^4
q}{(2\pi)^4}\left\{ \fr{k\!\!\!/ }{(k^2-m^2_b)}
\fr{1}{(k^2-m^2_\rho)}\fr{q\!\!/}{(q^2-m^2_c)}
\fr{1}{(q^2-m^2_\rho)}\right.\crn & &\times \left.
\fr{1}{[(k-q)^2-m^2_\rho]}\fr{1}{[(k-q)^2-m^2_\eta]} \right\}
\label{ctgg}
 \eea
 \ede
Note that the minus sign  in Eq. (\ref{ctgg}) is again due to
summation over the $SU(3)$ indexes.  Diagrams \ref{figdl3}(c) --
\ref{figdl3}(f) give a total contribution \bde \bea
M^{(G)}_{ai,dj}& = & M^{(G1)}_{ai,dj} + M^{(G2)}_{ai,dj} =
a_{ij}\sum_{b, c} \left[F_{ab} G_{bc} (G_{cd}^+) + F_{db} G_{bc}
(G_{ca}^+)\right] \crn & & \times  \int \fr{d^4 k}{(2\pi)^4}
\fr{d^4 q}{(2\pi)^4}\left\{ \fr{k\!\!\!\!/
}{(k^2-m^2_b)}\fr{q\!\!\!/ }{(q^2-m^2_c)} \fr{1}{(q^2-m^2_\rho)}
\fr{1}{(k^2-m^2_\rho)} \right.\crn & &\times \left. \left[
\fr{1}{(k^2-m^2_\eta)} \fr{1}{[(k-q)^2-m^2_\rho]} -
\fr{1}{[(k-q)^2-m^2_\eta]}
\fr{1}{[(k-q)^2-m^2_\rho]}\right]\right\} \label{contg2} \eea \ede

With the above mentioned approximation (that the integral being
relatively insensitive to $m_b,  m_c$) we have \bde \bea
M^{(G)}_{ai,dj} &\simeq & a_{ij} \sum_{b,c}\left[F_{ab} G_{bc}
(G_{cd}^+) + F_{db} G_{bc} (G_{ca}^+)\right]
 \left(\fr{1}{16 \pi^2}\right)^2
\fr{1}{{\overline M}^2} \label{contg21} \eea \ede

The contribution is again
proportional to $ (FGG^+)^T =$ $ G^* G^T F^T =$ $ - G^* G^T F$.
This shows that two-loop contribution as expected,
 is symmetric.  Note that only coupling constant ${\overline
\la}_{14}$ contributes.  In terms of the mass matrix, the G-type
contribution gives
 \bea M^{(G)}_{N}& \simeq &
p \sum_{b, c}\left[F_{ab} G_{bc} (G_{cd}^+) + F_{db} G_{bc}
(G_{ca}^+)\right], \label{contg3} \eea where \be p = n_2
\left(\fr{1}{16 \pi^2}\right)^2 \fr{1}{{\overline M}^2} \label{p}
\ee as before,  $ n_2 = 2 \mu_5 \bar{\la}_{14} \om^2$.
 Two-loop contributions to
$M_{\nu }$ have the similar forms with just $a_{ij}$ is replaced
by $ n_{ij} \equiv  + 2 \de_{1i} \de_{1j} {\overline \la}_{14}
\mu_5  v^2 $  and the mass of $\eta$ in propagator is replaced by
the mass of the $\chi$ boson. Note  the  plus sign on  $n_{ij}$
and in this case $n_{ij}$ is nonzero if $i = j = 1$. The VEV
$v_\eta$ corresponds to the first component in the $\eta$ triplet.

 Phenomenologically, it is necessary to fine-tune such that \be G_{ab} \gg
F_{ab}, \label{cond} \ee  because $G_{ab}$ is the charged lepton
mass, while $F_{ab}$ is the neutrino Dirac mass. However, such
fine-tuning is technically due to the protection $L_N$ symmetry as
discussed earlier.
 Therefore  the  F-type contributions in diagrams
 \ref{figdl3}(a) and \ref{figdl3}(b) are
negligible and,  hence $M_N \simeq M_N^{(G)}$ .

 To look at $M^{G}$ more closely , we can always choose a basis so
that $G_{a b}$ ( and $h_{2 \al \bet}$ in the case of quarks) is
{\it diagonal}. We have then \bea && \left(M^{(G)}_{N}\right)_{11}
= 2 p \sum_{b, c}F_{1b}
G_{bc} G^*_{1c} = 2p F_{11} G_{11}^2 = 0,\\
 && \left(M^{(G)}_{N}\right)_{12}= 2p \sum_{b, c} F_{1b}
G_{bc} G^*_{2c} \crn && = 2p  F_{12}\left( | G_{22}|^2 - |
G_{11}|^2\right) = \left(M^{(G)}_{N}\right)_{21}\nn \eea

Note that in the approximation (\ref{cond}), our result is similar
to the one-loop radiative corrections~\cite{cz}.
 Hence Eq.(\ref{matn}) becomes
\bde \bea M_{N}\simeq M^{(G)}_N \simeq  2 p \left(
\begin{array}{ccc} 0 & F_{12}\left( | G_{22}|^2 - |
G_{11}|^2\right)  & F_{13}\left( | G_{33}|^2 - | G_{11}|^2\right)
\crn F_{12}\left( | G_{22}|^2 - | G_{11}|^2\right)  & 0 &
F_{23}\left( | G_{33}|^2 - | G_{22}|^2\right) \crn F_{13}\left( |
G_{33}|^2 - | G_{11}|^2\right) & F_{23}\left( | G_{33}|^2 - |
G_{22}|^2\right) &0
\end{array} \right).
\label{mat3}\\
\eea \ede

 Noting  that  $G_{11} \sim m_e$, $G_{22} \sim m_\mu$, $G_{33} \sim
m_\tau$, then the matrix in Eq. (\ref{mat3}) has the form
 \bea M_{N}
\simeq 2 p \left(
\begin{array}{ccc} 0 & F_{12}\left( \fr{m_\mu}{\langle \rho_2
\rangle }\right)^2 & F_{13}\left( \fr{m_\tau}{\langle \rho_2
\rangle }\right)^2
 \\ F_{12}\left(
\fr{m_\mu}{\langle \rho_2 \rangle }\right)^2  & 0 & F_{23}\left(
\fr{m_\tau}{\langle \rho_2 \rangle }\right)^2 \\ F_{13}\left(
\fr{m_\tau}{\langle \rho_2 \rangle }\right)^2
 & F_{23}\left(
\fr{m_\tau}{\langle \rho_2 \rangle }\right)^2  &0
\end{array} \right)
\label{mat4}\eea

 Denoting  $A= \fr{u}{\sqrt{2}} F_{12}, B = \fr{u}{\sqrt{2}}  F_{13},
  C = \fr{u}{\sqrt{2}}  F_{23}$, then we can rewrite
  \bea M_{d} \simeq \left(
\begin{array}{ccc} 0 & A & B
 \\ - A & 0 & C \\ -B
 & -C  &0
\end{array} \right)
\label{mat41}  \eea
 Note that the relative size of $M_N$ relative to $M_d$ is
controlled by the scale ratio $(\omega/\overline{M})^2$ times some
two-loop factor.  As we states before, if the ratio
$(\omega/\overline{M})^2$ is chosen to be very large so as to
overcome the two-loop suppression factor, the usual seesaw
scenario can still apply. However, here we shall concentrate on
the more interesting, and probably more natural, case when
$(\omega/\overline{M})^2$ times some two-loop factor is small such
that $M_N$ can be considered a small perturbation to the Dirac
mass $M_d$.
 Next  denoting $r = \fr{m_\mu^2}{m_\tau^2} \ll 1$,
 $d\equiv  2 p\ F_{12} \left(\fr{m_\mu}{\langle
 \rho_2 \rangle }\right)^2$ $ = \fr{4
 \sqrt{2}}{u^3} p r A  m^2_\tau, s\equiv
 \fr{4
 \sqrt{2}}{u^3} p B  m^2_\tau,
 t\equiv
 \fr{4
 \sqrt{2}}{u^3} p C  m^2_\tau
$ \  then the neutrino mass matrix has the form
 \bea M_{\nu N}
=\left(
\begin{array}{cccccc}
 &&& 0& A& B
 \\
 & {\cal O}  & & - A&0& C
 \\
  & & &- B&- C& 0\\
0 &-A&-B&0 &d &s \\
 A&0&-C&d  & 0& t    \\
B&C &0 & s&t &0   \\
\end{array}
\right) \label{mat7}
 \eea
 with  $d \ll s, t \ll A, B, C$.
 We assume that $u \simeq v $ so  $u = \fr{v_{SM}}{\sqrt{2}}
\simeq  175 $  GeV.

\section{Phenomenology}
\label{pheno} The interesting new physics compared with other
3-3-1 models is the neutrino physics. By our convention and from
(\ref{mat41}), it follows  that $M_d$  is anti-Hermitian $M_d^+ =
- M_d$, therefore its eigenvalues are imaginary  and given $0, \pm
i L, \ L \equiv \sqrt{A^2+B^2+C^2}$. The eigenstates are
 \bea  \left(
\begin{array}{c} \nu_1\\
\nu_2\\ \nu_3
\end{array} \right) = U_3^+ \left(
\begin{array}{c} \nu_e\\
\nu_\mu\\ \nu_\tau
\end{array} \right)
\label{matv} \eea where \bde
 \bea U_3 = \left(
\begin{array}{ccc}
\fr C L  &\fr{1}{L'} (  B C - i A L)& \fr{1}{L'} (  B C + i A L)
 \\
-\fr B L  &  \fr{1}{L'} ( A^2 +  C^2)&\fr{1}{L'} ( A^2 +  C^2)
 \\
\fr A L  &\fr{1}{L'} (   AB + i C L )& \fr{1}{L'} ( A B - i C L)
\end{array} \right)
=\left( \vec{\phi_0},   \vec{\phi_+},  \vec{\phi_-}
 \right),
\label{matu} \eea \ede where $L' = L \sqrt{2 (A^2+C^2)}$ and
$\phi_i$ are normalized eigenvectors of $M_d$. The unitary matrix
$U_3$ diagonalizes $M_d$, i.e.
 \bea
U_3 U_3^\dagger & = & U_3^\dagger U_3 = I, \crn
 U_3^\dagger M_d U_3 & = & D_d = {\rm diag} ( 0, -  i L, i L)
 \label{umat}
  \eea

We see that in the tree level we have three Dirac eigenstates. Two
of them have degenerate eigenvalues $L$ and the other one
massless.  It is easy to identify the mass splitting $L$ as the
value of measured atmospheric neutrino mass difference $\Delta
m_{atm}$.  Therefore we require the parameter $A, B, C$ to be of
order $\Delta m_{atm} \sim 5 \times 10^{-2}$ ~eV which is much
smaller the charged lepton mass.  This is of course part of the
fine-tuning in fermion Yukawa couplings we need in this model. At
loop level, this inverted spectrum is corrected by $M_N$, it will
not only give rise to mass splitting between the two degenerate
Dirac states, it will also split each Dirac pairs into two
non-degenerate Majorana states, resulting in the spectrum with six
Majorana eigenstates with four heavier ones and one light one and
one remains massless.   The existence of the  massless Majorana
state is a result of our approximation which gives $M_{\nu N}$
with zero determinant at the level of our approximation.   We
expect all the smaller (Majorana) mass splitting due to $M_N$
should be of the order of $\Delta m = \Delta m^2_{sol}/ \Delta
m_{atm} \sim 8 \times 10^{-4}$~ eV.  Here we are assuming that the
solar oscillation  is between the two heavier Majorana states. So
for $A, B, C$ of the same order of magnitude, $5 \times 10^{-2}$
~eV, we expect $s, t$ to be of order $8 \times 10^{-4}$~ eV.

More specifically, with our loop correction $d, s, t \neq 0$, we
can take them as perturbation and diagonalize the $6\times6$ mass
matrix.  (See the appendix  for more details). Note that, if we
ignore CP violation, the mass matrix $M_{\nu N}$ is real,
symmetric and therefore Hermitian.  It can be diagonalized by an
unitary matrix with real eigenvalues. When $d=s=t=0$, the
eigenvalues of $M_{\nu N}$ are $L, 0, -L$ and eigenvectors of
$M_{\nu N}$ are \bea \Psi_1^T = ( \vec{\phi_+} , & - i \vec\phi_+
)/\sqrt{2},  \crn \Psi_2^T = ( \vec\phi_- , & i \vec\phi_-
)/\sqrt{2} \eea for eigenvalue $L$; \bea \Psi_5^T = ( \vec{\phi_+}
,  & i \vec\phi_+ )/\sqrt{2},  \crn \Psi_6^T = ( \vec\phi_- ,  & -
i \vec\phi_- )/\sqrt{2} \eea for eigenvalue $-L$;
 and \bea
\Psi_3^T = ( \vec{\phi_0} ,  & 0 ),
 \crn
\Psi_4^T = ( 0 ,  &  \vec\phi_0 ) \eea for eigenvalue $0$. In this
basis, we can use $M_N$ as perturbation and calculation the lowest
order correction of $d, s, t$ to the $6\times6$ mass matrix.  The
result (from the appendix) can be written as \bea \Delta M_{\nu N}
= \left(
\begin{array}{ccc}
 \Delta_L  &  &
 \\
& \Delta_0&
 \\
& & \Delta_{-L}
\end{array} \right)
\label{deltam} \eea where $\Delta_i$ are the diagonal $2\times2$
blocks. Here we present only the diagonal blocks because they are
the states with degenerate eigenvalues and therefore give the
leading order corrections. The other off-diagonal blocks are
nonzero, (they will be given in the appendix), but they only
contribute at higher order.  The $\Delta_i$ are \bea \Delta_L =
\Delta_{-L} = \left(
\begin{array}{cc}
 \Delta_{++}/2  &  - \Delta_{+-}/2
 \\
- \Delta_{-+}/2& \Delta_{++}/2
\end{array} \right)
\label{deltam} \eea and \bea \Delta_0 = \left(
\begin{array}{cc}
 0  &  0
 \\
0 & \Delta_{00}
\end{array} \right)
\label{deltamo} \eea
 where $\Delta_{ij}$ are given by
\bde
  \bea
 \De_{++} & = & \fr{1}{L^2} (B C d  - A C s + A B t),\label{dec}\\
\De_{-+}&=& \left(\fr{ BC - i AL}{L^2}\right) d + \left[\fr{AC(A^2
+ 2 B^2+C^2) + i B(C^2 - A^2)L}{(A^2+C^2)L^2}\right]s\crn
& & + \left(\fr{AB + i CL}{L^2}\right)t,\label{dectt}\\
 \De_{00} &=& - 2 \De_{++}
 \label{del}
 \eea
 \ede
They satisfy the properties $\De_{++} = \De_{--}, \hs \De_{0-} =
\De_{0+}^*$, and $\De_{ij} = \De_{ji}^*$.
 The eigenvalues of matrix $M_{\nu N}$ are given by
 \bea m_{1,2}& =& L +
\fr{\De_{++}}{2} \pm \fr 1 2 \sqrt{|\De_{+-}|^2},\\
 m_{5,6}& =&  - L +
\fr{\De_{++}}{2} \pm \fr 1 2 \sqrt{|\De_{+-}|^2},\nn \eea \bea
m_3&=&0,\\
 m_4 &=& -2 \De_{++}.
 \label{rootmn}
 \eea
Note that $\De_{+-}$ is complex and \bde
 \bea
 |\De_{+-}|^2 & = & \fr{1}{L^4 (A^2 + C^2)^2}\left\{(A^2 + C^2)^2 [ (B^2
 C^2 + A^2 L^2) d^2  + (A^2
 B^2 + C^2 L^2) t^2\right.\crn
 & & - 2 (A^2 + C^2) A C dt] + (A^2 B^2 + C^2 L^2)(B^2
 C^2 + A^2 L^2) s^2\crn
 & & + \left.2 (A^2 + C^2) [ (B^2 C^2 + A^2 L^2) A B d s +
 (A^2 B^2 + C^2 L^2)B C s t]\right\}.
\label{dect} \eea
 Therefore $\sqrt{|\De_{+-}|^2}$ is nonanalytic
in $d, s, t$.
 For example, in the simplified case of $A=B=C={\cal M}$, we have

 $ L= {\cal M} \sqrt{3},$
 $\De_{++} = \fr 1 3 (d-s+t)$,
 $|\De_{+-}|^2  =  \fr{4}{3^2} (d^2 +s^2+t^2 + d s + s t - d
 t)$.\ede
In this case, the eigenmasses are given \bde \bea m_{1,2}& =&
{\cal M} \sqrt{3}
+ \fr 1 6 (d-s+t) \pm \fr 1 3 \sqrt{(d^2 +s^2+t^2 + d s + s t - d t)},\\
 m_{5,6}& =& - {\cal M} \sqrt{3}
+ \fr 1 6 (d-s+t) \pm \fr 1 3 \sqrt{(d^2 +s^2+t^2 + d s + s t - d
t)},\label{eigabc2}\eea \ede
\bea
m_3&=&0,\\
 m_4 &=& - \fr 2 3 (d-s+t),
 \label{eigabc}
 \eea

 The eigenvectors of $m=0$ and $m=\Delta_{00}$
 have a form \bea  && \fr{1}{\sqrt{2}}(\Psi_3+ i\Psi_4),\crn &&
      \fr{1}{\sqrt{2}}(\Psi_3- i\Psi_4)
      \eea
For other eigenvectors with eigenvalues $L$, we get \bea
\fr{1}{\sqrt{2}} \left(
\begin{array}{c} 1\crn
 e^{- i \phi }
\end{array} \right)& =& \fr{1}{\sqrt{2}}(\Psi_1  +
e^{ - i \phi }\Psi_2
),\\
 \fr{1}{\sqrt{2}} \left(
\begin{array}{c} 1\crn
- e^{ i \phi }
\end{array} \right) &=&  \fr{1}{\sqrt{2}}(\Psi_1  -
e^{ - i \phi } \Psi_2 ) \eea where $\phi$ is the phase of
$\De_{+-}= |\De_{+-}|e^{i\phi}$. The eigenvectors with eigenvalue
$-L$ are similar linear combination of  $\Psi_5 $ and
 $\Psi_6 $.

The mixing matrix in the basis $\nu_e, \nu_\mu,\nu_\tau, N_1, N_2,
N_3$ and $\nu_1, \nu_2, \nu_3, \nu_4, \nu_5, \nu_6$ is given by
\bea  \left(
\begin{array}{c} \nu_e\\
\nu_\mu\\ \nu_\tau\\ N_1\\ N_2\\ N_3
\end{array} \right) = U\left(
\begin{array}{c} \nu_1\\
\nu_1\\ \nu_3\\ \nu_4\\ \nu_5\\ \nu_6
\end{array} \right)
\label{matv}
 \eea
where \bde
 \bea && U= \fr{1}{2 L'}\left(
\begin{array}{cccccc}
 U_{e 1}  & U_{e 2}  & \fr{C L'\sqrt{2}}{L} &
\fr{C L'\sqrt{2}}{L}&  U_{e 1} &  U_{e 2}
 \\
  U_{\mu 1}  & U_{\mu 2}  & -\fr{B L'\sqrt{2}}{L} &
-\fr{B L'\sqrt{2}}{L}&  U_{\mu 1} &  U_{\mu 2}
 \\
   U_{\tau 1}  & U_{\tau 2}  & \fr{A L'\sqrt{2}}{L} &
\fr{A L'\sqrt{2}}{L}&  U_{\tau 1} &  U_{\tau 2}
 \\
    U_{N_1 1}  & U_{N_1 2}  & \fr{iC L'\sqrt{2}}{L} &
\fr{-iC L'\sqrt{2}}{L}& - U_{N_1 1} & - U_{N_1 2}
 \\
     U_{N_2 1}  & U_{N_2 2}  & -i\fr{B L'\sqrt{2}}{L} &
i\fr{B L'\sqrt{2}}{L}& - U_{N_2 1} & - U_{N_2 2}
 \\
      U_{N_3 1}  & U_{N_3 2}  & \fr{iA L'\sqrt{2}}{L} &
-\fr{iA L'\sqrt{2}}{L}& - U_{N_3 1} & - U_{N_3 2}
\end{array} \right)
\label{matsh}  \eea \ede
 where the matrix elements are given in
  Table \ref{uelement}.

\begin{table}
\caption{
     Matrix elements}
\begin{center}
\begin{tabular}{c|c}
\hline
 $U_{e 1}$ & $BC\ka_+ - iAL \ka_-$\\
$U_{e2}$&$  BC\ka_- - iAL \ka_+$ \\
$U_{\mu 1}$&$  (A^2+C^2)\ka_+$ \\
$U_{\mu 2}$&$ (A^2+C^2)\ka_-$ \\
 $U_{\tau 1}$&$  AB\ka_+ + iCL\ka_- $\\
$U_{\tau 2}$&$  AB\ka_- + iCL\ka_+ $\\
$U_{N_ 1 1}$&$ -AL\ka_+ - iBC\ka_-$\\
$U_{N_ 1 2}$&$ -AL\ka_- - iBC\ka_+$\\
$U_{N_2 1}$&$ -i(A^2+C^2)\ka_-$\\
 $U_{N_2 2}$&$ -i(A^2+C^2)\ka_+$\\
 $U_{N_3 1}$&$ CL\ka_+ - iAB\ka_-$\\
  $U_{N_3 2}$&$ C L\ka_- - i AB\ka_+$.\\
\end{tabular}
\label{uelement}
\end{center}
\end{table}
 Here we have denoted  $k_{\pm} = 1  \pm e^{-i\phi}$.
Note that $U_{N_1 1}= - i U_{e2}$, $U_{N_2 1}= - i U_{\mu2}$,
$U_{N_3 1}= - i U_{\tau2}$, $U_{N_1 2}= - i U_{e1}$, $U_{N_2 2}= -
i U_{\mu1}$, $U_{N_3 2}= - i U_{\tau1}$ and $ \fr{L'\sqrt{2}}{L} =
2 \sqrt{A^2 + C^2}$.

One can check out that this matrix is unitary
 \be  U U^\dagger= U^\dagger U =  I.
\label{unity}  \ee

Here the inverted hierarchy neutrino mass spectrum  is used and is
shown in Fig.\ref{nspec}. \vs

\begin{figure}[htbp]
\begin{center}
\includegraphics[width=4cm,height=4cm]{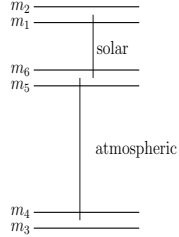}
\caption{\label{nspec}{The  inverted hierarchy neutrino mass
spectrum, showing the usual solar and atmospheric mass differences
}}
\end{center}
\end{figure}

 The survival probability is given by \cite{mohapal}
  (in the extremely relativistic limit)
  \bea  P_{l' l} & =& \vert \langle \nu_{l'}(0) | \nu_l \rangle(x) \vert^2 =
\sum_{\al,\bet}\vert U_{l\al} U_{l'\al}^* U_{l\bet}^*
U_{l'\bet}\vert \crn && \times \cos \left( \fr{2\pi x}{L_{\al
\bet}} - \vph_{ll'\al \bet}\right), \label{tran} \eea where
$L_{\al \bet} \equiv \fr{4 \pi |{\bf p}|}{\De m^2_{\al \bet}}$ and
$\vph_{ll'\al \bet}$ is the phase of $U_{l\al} U_{l'\al}^*
U_{l\bet}^* U_{l'\bet}$ , with  $ \De m^2_{\al \bet} = m_\al^2 -
m_\bet^2$.

In the literature, usually the only simplest two component
neutrino mixing result is used in the analysis. However, in our
case of six light Majorana neutrinos, the analysis is much more
complicated  and has not been fully explored in the literature.
Even the oscillation formula for three flavor which is available
in the literature is far from useful here.

For our special neutrino spectrum, however, we can make some
approximation and get some nice result.  Since the small $\Delta
m^2_{sol}$ is mainly between $\nu_e$ and $\nu_\mu$ and the larger
$\Delta m^2_{atm}$ is mainly between $\nu_\tau$ and $\nu_\mu$, it
is reasonable to assume that  $\nu_\tau$ is mainly contained in
$\nu_3$ and $\nu_4$ which are very light, while $\nu_e$ and
$\nu_\mu$ are mainly contained in $\nu_1$, $\nu_2$, $\nu_5$,
$\nu_6$ which are almost degenerate in mass.  In any case, this is
almost the only reasonable guess for any theory with an inverted
hierarchy neutrino mass spectrum.  We shall assume this here.

In this case we can get an approximate result for $P_{\mu\tau}$ in
the vacuum, by considering only the transition between light mass
eigenstates
  $\nu_3$ and $\nu_4$ and those of heavier $\nu_1$, $\nu_2$, $\nu_5$,
  $\nu_6$.
Since the other transitions involving states whose mass splitting
are too small for atmospheric range oscillation. Therefore
%The other transitions involving states with mass splitting are too
%small for atmospheric range oscillation. Therefore
 \bea P_{\mu
\tau} & \approx & \sum_{\al = 3, 4} U_{\mu\al} U_{\tau\al}^*
U_{\mu\al}^* U_{\tau\al} + \sum_{\bet = 1, 2, 5, 6} U_{\mu\bet}
U_{\tau\bet}^* U_{\mu\bet}^* U_{\tau\bet}
 \crn
& & + \sum_{\al = 3, 4} \sum_{\bet = 1, 2, 5, 6} | U_{\mu\al}
U_{\tau\al}^* U_{\mu\bet}^* U_{\tau\bet}|\crn && \times
 \cos
\left( \fr{2\pi x}{L_{\al \bet}} - \vph_{\mu\tau\al \bet}\right),
\label{tranmt}
 \eea
 where $U$ is given in Eq.(\ref{matsh}), for
example, $U_{\mu 1} = U_{\mu 5} = \fr{ (A^2+C^2)\ka_+}{ 2 L'}$,
while $U_{\tau 3} = \fr{ A}{L\sqrt{2}}$. $x$ is in the range of
atmospheric neutrino oscillation \cite{mohapal}.

 Substituting matrix elements into Eq. (\ref{tranmt}) we see
that the first term in (\ref{tranmt}) is given by
 \bea  \sum_{\al
= 3, 4} U_{\mu\al} U_{\tau\al}^* U_{\mu\al}^* U_{\tau\al} =  \fr{
A^2 B^2}{ 2 L^4},\label{mt1}\eea and the second term in
(\ref{tranmt})   has the form
 \bea && \sum_{\bet = 1, 2, 5, 6 }U_{\mu\bet} U_{\tau\bet}^*
 U_{\mu\bet}^* U_{\tau\bet}
 = \fr{1}{4 L^4}[ A^2B^2+C^2L^2\crn && + \cos^2 \phi (A^2B^2-C^2L^2)
 - A B C L \sin 2\phi]
 \label{sumb6} \eea

Let us calculate the third term
 \bea  \sum_{\al
= 3, 4} && U_{\mu \al} U_{\tau \al}^* = - \fr{AB}{
 L^2},\label{sumb6}\\
 \sum_{\bet = 1,2, 5,6}&& U_{\mu \bet}^*
U_{\tau \bet} =  \fr{8 AB (A^2+C^2)}{(2 L')^2} = \fr{ AB }{
L^{2}}. \label{surtp} \eea
 Here two factors are all real.
Therefore $\vph_{\mu\tau\al \bet}$ can be ignored.
 Hence
\bde
 \bea   &&
 \sum_{\al = 3, 4} \sum_{\bet = 1, 2, 5, 6} | U_{\mu\al}
U_{\tau\al}^* U_{\mu\bet}^* U_{\tau\bet}|
 \cos
\left( \fr{2\pi x}{L_{\al \bet}} - \vph_{\mu\tau\al
\bet}\right)
%\crn &=&
=  - \fr{ A^2 B^2 }{ L^4} \cos \left( \fr{2\pi
x}{L_{atm}}\right).\label{su7} \eea
 We get then transition probability
 \bea P_{\mu \tau}& = & \fr{1}{4 L^4}\left[
3 A^2 B^2 + C^2 L^2  + \cos^2\phi ( A^2 B^2 - C^2 L^2 ) - A B C
\sin 2\phi\right.
%\crn &  &
- \left. 4 A^2 B^2 \cos \left( \fr{2\pi
x}{L_{atm}}\right) \right],\label{surmt}
 \eea
 \ede
  where
\be L^{-1}_{atm} = \fr{ ( m^2_{1,2,5,6}- m^2_{3,4}) }{4 \pi p}
\simeq \fr{L^2 } {4 \pi p}. \label{lat}\ee
 For $\nu_e ,\nu_\tau$ oscillation
we have \bde
 \bea P_{e \tau} & \approx & \sum_{\al =
3, 4} U_{e\al} U_{\tau\al}^* U_{e\al}^* U_{\tau\al} + \sum_{\bet =
1, 2, 5, 6} U_{e\bet} U_{\tau\bet}^* U_{e\bet}^* U_{\tau\bet}
%\crn & &
 + \sum_{\al = 3, 4} \sum_{\bet = 1, 2, 5, 6} | U_{e\al}
U_{\tau\al}^* U_{e\bet}^* U_{\tau\bet}|  \cos \left( \fr{2\pi
x}{L_{\al \bet}} - \vph_{e\tau\al \bet}\right) \label{tranet}
 \eea
\ede Using a similar approximation, we obtain \bde
 \bea
\sum_{\al = 3, 4}&&  U_{e\al} U_{\tau\al}^* U_{e\al}^* U_{\tau\al}
=\fr{32 A^2 C^2 (A^2+C^2)^2}{(2 L')^4},\crn \sum_{\bet = 1, 2, 5,
6}&& U_{e\bet} U_{\tau\bet}^* U_{e\bet}^* U_{\tau\bet}
 = \fr{1}{(2L')^4}\left\{8[B^2 C^2 + A^2 L^2 +
\cos^2 \phi(B^2 C^2 - A^2 L^2) \right. \crn &&+\left. 2 A B C L
\sin \phi\right][ A^2 B^2 + C^2 L^2  + \cos \phi (A^2 B^2 - C^2
L^2 ) - 2 ABC L \sin \phi]
 \crn
 &&+8[B^2 C^2 + A^2 L^2 -
\cos^2 \phi(B^2 C^2 - A^2 L^2) \crn &&- \left. 2 A B C L \sin
\phi][ A^2 B^2 + C^2 L^2  - \cos \phi (A^2 B^2 - C^2 L^2 ) +  2
ABC L \sin \phi] \right\},\label{surtp34}\eea \ede
 \bea
 \sum_{\al = 3, 4}&& U_{e \al} U_{\tau \al}^* = \fr{AC}{
L^2},\crn \sum_{\bet = 1,2, 5,6}&& U_{e \bet}^* U_{\tau \bet} =
-\fr{2AC (A^2+C^2)}{ L^{'2}}. \label{surtp} \eea
%\end{widetext}
These values again are real. Summing up, we get \bde
 \bea
P_{e \tau} & = & \fr{1}{4(A^2+C^2)^2 L^4} \left\{2 A^2 C^2
(A^2+C^2)^2 + ( A^2 B^2 + C^2 L^2 )(B^2 C^2 + A^2 L^2)\right. \crn
&&+( A^2 B^2 - C^2 L^2 )[\cos^2 \phi (B^2 C^2 - A^2 L^2) + A B C L
\sin 2 \phi]\crn &&-\left. A B C L [(B^2 C^2 - A^2 L^2)\sin 2 \phi
+ 4 A B C L \sin^2 \phi]\right\} -
 \fr{ A^2 C^2}{ L^4 } \cos \left( \fr{2\pi
x}{L_{atm}}\right) \label{surett} \eea \ede
 From (\ref{surett}) we see
that to get    $\sin^2 \theta_{13} \approx 0$ we just need $C=0,
\phi = 0 $ or equivalently \be C=0, \  d = s = 0.
 \label{condpet} \ee
In this limit, the atmospheric neutrino transition probability
(\ref{surmt}) becomes \bea P_{\mu \tau}& = & \fr{A^2 B^2 }{
L^4}\left[ 1 -   \cos \left( \fr{2\pi x}{L_{atm}}\right) \right]=
2\fr{A^2 B^2 }{ L^4} \crn & & \times \sin^2\left( \fr{\pi
x}{L_{atm}}\right) \label{conzero} \eea
 Eq.(\ref{conzero}) gets the usual
form~\cite{bil}
 \bea P_{\mu \tau}& = & \fr{1}{2}  \sin^2 2\theta_{atm}
  \sin^2\left( \fr{\pi x}{L_{atm}}\right)
  \label{three} \eea
 by identification
 \be
    \sin^2 2\theta_{atm} = 4  \fr{A^2 B^2 }{ L^4}  = \fr{4A^2
    B^2}{(A^2+B^2)^2}
 \label{cons} \ee
 Thus
 \be
    \sin^2 2\theta_{atm} =  1 \  \Rightarrow \   A = B.
 \label{ab} \ee
 To finish this step, we note that to get $\sin^2
\theta_{13} \approx 0$ and $\sin^2 2\theta_{atm} \simeq 1$, one
just need $C=0,  \  A = B$.

 It is much harder to make an approximate calculation for $P_{e
\mu}$ since it involves \bea P_{e \mu} & \approx & \sum_{\alpha
\bet = 1, 2, 5, 6} | U_{e\al} U_{\mu\al}^* U_{e\bet}^*
U_{\mu\bet}|
 \cos
\left( \fr{2\pi x}{L_{\al \bet}} - \vph_{e\mu\al \bet}\right),\nn
\\
\label{tranetn} \eea for $x$ is in the range of solar neutrino
oscillation.  It is hard to proceed analytically. However, we can
make a numerical study of the possibility to make $\sin^2
\theta_{sol} \simeq 0.3$. Of course, the solar neutrino
oscillation is mostly likely due to matter-induced MSW oscillation
and our vacuum oscillation treatment is flawed, but it serve to
illustrate that our model can easily fit the solar data also.  A
more detailed serious study of the six Majorana flavor (for this
or other similar pseudo-Dirac models~\cite{pseudo})  will be
needed. This and a study of the astrophysics constraint will be
investigated in a future publication.

Now we consider $\nu_e, \nu_\mu$ transition and for the sake of
shorthand denote $q \equiv \sqrt{|\De_{+-}|^2}$.  We have \bea
m_1^2&=&L^2+ \fr{\De_{++}^2}{4} +\fr{q^2}{4} + L\De_{++} + Lq +
\fr 1 2 \De_{++} q,\crn
 m_2^2&=&L^2+ \fr{\De_{++}^2}{4}
+\fr{q^2}{4} + L\De_{++} - Lq -\fr 1 2 \De_{++} q,\crn
m_5^2&=&L^2+ \fr{\De_{++}^2}{4} +\fr{q^2}{4} - L\De_{++} - Lq +
\fr 1 2 \De_{++} q,\crn
 m_6^2&=&L^2+ \fr{\De_{++}^2}{4}
+\fr{q^2}{4} - L\De_{++} + Lq - \fr 1 2 \De_{++} q\label{massdif}
\eea Some manipulations  give \bde
 \bea  P_{e \mu}(A,B,C,d,s,t)  &
= & \fr{ \cos \phi}{4 L^4} \left[(B^2C^2- A^2 L^2)\cos \phi + 2
ABC L \sin \phi\right]\crn & &+ \fr{1}{8 L^4}\left \{ \cos \phi
\left[(B^2C^2- A^2 L^2)\cos \phi + 2 ABC L \sin \phi\right]\right.
\crn && \times
 \left(
\cos\left[\fr{L(\De_{++} + q)}{ {\bar{m}^2}}\right] +
\cos\left[\fr{L(\De_{++} - q)}{ {\bar{m}^2}}\right]\right) \crn &&
+ \sin \phi \left[(B^2C^2- A^2 L^2)\sin \phi - 2 ABC L \cos
\phi\right]\crn &&\times \left(\cos\left[\fr{q(2L + \De_{++})}{
{2\bar{m}^2}}\right] + \cos\left[\fr{\De_{++} (2L + q)}{2
{\bar{m}^2}}\right]\right.\crn &&+\left.\cos\left[\fr{\De_{++} (2L
- q)}{2 {\bar{m}^2}}\right] +\left. \cos\left[\fr{q(2L -
\De_{++})}{ {2\bar{m}^2}}\right]\right) \right\}
 \label{tranemq} \eea
 \ede
where \be \bar{m}^2 = \fr{|{\bf p}|}{x} = 10^{-11}
[eV^2]\label{leng}\ee is the solar oscillation parameter.

In the case  $ C=0$,
 the $\nu_e \nu_\mu$ transition probability becomes
\bde
 \bea  P_{e \mu}(A,B,0,d,s,t) & = & \fr{A^2}{4 L^2} \cos^2
\phi\crn & & + \fr{A^2}{8 L^2}\left \{ \cos^2 \phi
 \left(
\cos\left[\fr{L(\De_{++} + q)}{ {\bar{m}^2}}\right] +
\cos\left[\fr{L(\De_{++} - q)}{ {\bar{m}^2}}\right]\right)\right.
\crn && + \sin^2 \phi  \left(\cos\left[\fr{\De_{++} (2L + q)}{2
{\bar{m}^2}}\right]+\cos\left[\fr{\De_{++} (2L - q)}{2
{\bar{m}^2}}\right]\right.\crn && +\cos\left[\fr{q(2L +
\De_{++})}{ {2\bar{m}^2}}\right]   +\left. \left.
\cos\left[\fr{q(2L - \De_{++})}{ {2\bar{m}^2}}\right]\right)
\right\}
 \label{tranemq} \eea
 \ede
Putting one more condition, $  d=0$ (together with $C=0$) we have
$L=\sqrt{A^2+B^2}$ and \bea \cos \phi & = & \fr{At}{\sqrt{
s^2(A^2+B^2) + A^2t^2}},\crn  \sin \phi& =& \fr{s
\sqrt{A^2+B^2}}{\sqrt{ s^2(A^2+B^2) + A^2t^2}},\eea \bea q & = &
\fr{B\sqrt{ s^2(A^2+B^2) + A^2t^2}}{A^2+B^2} , \crn \De_{++}& =&
\fr{ABt}{A^2+B^2}. \label{n3}\eea
 Substituting Eq.(\ref{leng}) and
(\ref{n3}) into (\ref{tranemq}) we get for $s=0$ \bde
 \bea P_{e
\mu}(A,B,0,0,0,t) & = & \fr{A^2}{4(A^2+B^2)} +
\fr{A^2}{8(A^2+B^2)}\left\{ 1 +  \cos \left[ 2 \times 10^{11}
\fr{A B t}{\sqrt{A^2+B^2}} \right]\right\}\nn\\
 \label{tranemt10} \eea
 \ede
In Fig. \ref{figt} we plotted $P_{e \mu}(0.12,0.1,0,0,0,t[0.1
\textrm{eV}])$ for $t$ run from 0 to $10^{-8}$ or from $0$ to
$10^{-9}\ \textrm{eV}$

\begin{figure*}[htbp]
\begin{center}
\includegraphics[width=10cm,height=7cm]{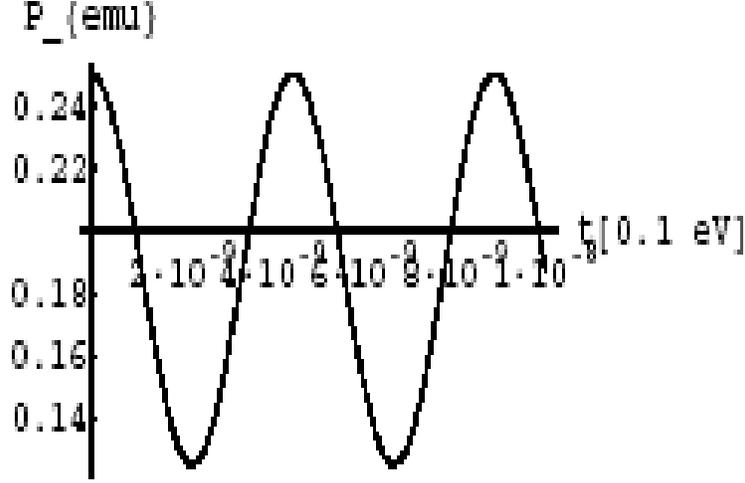}
\caption{\label{figt}{\rm $P_{e \mu}$  as function of $t$ within
$A =  B = 0.1\  \textrm{eV} , C = d = s = 0$. }}
\end{center}
\end{figure*}

The figure shows that, if $A$ and $B$  are  taken in order 0.1 eV
(upper cosmological bound) we have solar neutrino data for  $ t$
parameter.

 For the $\nu_e \nu_\mu$ transition in  matter, ignoring  magnetic
 field, we get typical terms similar to that in Ref.\cite{kl}.
 For general case,
$P_{\al \bet}$ cannot be cast in compact form and may require
numerical evaluation. Analysis of the matter effect is quite
simple in the case of two generations and gives a result in good
agreement with the current evaluation.

\section{Summary and  Conclusions}

The basic motivation of this work is to study neutrino mass and
mixing in the framework of the model based on
$\mbox{SU}(3)_C\otimes \mbox{SU}(3)_L \otimes \mbox{U}(1)_X$ gauge
group with right-handed neutrinos.

The Higgs sector of this model contains the bilepton Higgs scalars
with lepton number  $L=2$. Hence, the Yukawa coupling of the model
has automatic lepton number symmetry which is broken only by the
self-couplings of the Higgs boson. The interesting radiative
mechanism for neutrino masses has been obtained.  At the tree
level, the neutrino spectrum contains three Dirac fermions, one
massless and two degenerate in mass.  At the two-loop level,
neutrinos obtain Majorana masses and correct the tree-level result
which naturally gives rise  to the pseudo-Dirac mass differences
and an inverted hierarchy mass pattern  and interesting mixing
which can fit the current data with minor fine-tuning.

For the solar neutrino oscillations in matter, neglecting magnetic
field,  we have got the matter effects. Our analysis in the
simpler case limited by two generations shows that the scheme
gives appropriate consistency.

 In another scenarios, one can pick the scales
such that the loop-induced Majorana mass matrix is bigger than the
Dirac one and thus reproduce the usual seesaw mechanism.

The complete analysis and   study of the astrophysics constraint,
the medium effects  will be investigated in the future works.

\section*{Acknowledgement}
  H. N. L would like to thank the staff of the National Center
for Theoretical Sciences, where this work was done, for its
hospitality. He is grateful to the National Center for Theoretical
Sciences of the National Science Council (NSC) of the Republic of
China for financial support.  We both like to thank Prof. Hsiu-Hau Lin
for extensive discussion on the diagonalization of neutrino mass matrix.

\appendix

\section{Diagonalization of the mass matrix}

Let us consider a free case \bea H_0 = \left(
\begin{array}{cc}
 0  &  M
 \\
M^T & 0
\end{array} \right),
\label{lin1} \eea where $ M^T = - M,\hs  M^* = M $  and
 \bea M = \left(
\begin{array}{ccc} 0 & A & B
 \\ - A & 0 & C \\ -B
 & -C  &0
\end{array} \right)
\label{lin2}  \eea The interaction is considered as perturbation $
V = \left(
\begin{array}{cc}
 0  &  0
 \\
0 & \De
\end{array} \right),
$
with $ \De = \left(
\begin{array}{ccc} 0 & d & s
 \\ d & 0 & t \\ s
 & t  &0
\end{array} \right).
$ Let us start with a simple case $\De = 0$ then  $ H= H_0 + V =
\left(
\begin{array}{cc}
 0  &  M
 \\
-M & 0
\end{array} \right).
$
We search  eigenvectors  in the form $ \Psi_E = \left(
\begin{array}{c} \vph_A\crn
\vph_B
\end{array} \right)
$ satisfied the equation $ H\Psi_E = E\Psi_E $. Then we obtain an
equation \bea  \left(
\begin{array}{cc}
 0  &  M
 \\
-M & 0
\end{array} \right)
\left(
\begin{array}{c} \vph_A\crn
\vph_B
\end{array} \right) = E \left(
\begin{array}{c} \vph_A\\
\vph_B
\end{array} \right)
\label{lin7} \eea Equivalently, \bea M \vph_B & = & E\vph_A ,\crn
 - M \vph_A & =& E\vph_B.
\label{lin8} \eea From Eq. (\ref{lin8}), we get $ - M^2 \vph_A = E
M \vph_B = E^2 \vph_A,$ or $ (-M^2) \vph_A = E^2 \vph_A $. This
means that we need {\it only to diagonalize $M$}. Next, we
consider  a characteristic equation \bea \left(
\begin{array}{ccc} -\la  & A & B
 \\ - A & -\la & C \\ -B
 & -C  &-\la
\end{array} \right) =0
\label{lin11}  \eea
 or $ \la
(\la^2 + A^2 + B^2 + C^2) = 0.$
 Thus, we get three roots : $\la = 0,\  \pm i \sqrt{A^2 + B^2 +
 C^2}$.
 Let us denote
 $L = \sqrt{A^2 + B^2 + C^2}$, then
 \be E^2 = - \la_i^2 = 0, \  - (\pm i L)^2 = 0,\  L^2,\hs  L^2
  \label{lin13} \ee
  Thus we have three eigenvalues : $ + i L,\  0, \ - i L$.
  Let us choose $\vph_A$ to be $M's$ eigenstate
 \bea \left(
\begin{array}{ccc} 0 & A & B
 \\ - A & 0 & C \\ -B
 & -C  &0
\end{array} \right)
 \left(
\begin{array}{c} x\\
y\\ z
\end{array} \right) = 0
\label{lin14}  \eea It is easily to get $ \left(
\begin{array}{c} x\\
y\\ z
\end{array} \right) = \left(
\begin{array}{c} -C\\
B\\ -A
\end{array} \right).$
Let us change a sign of the eigenstate $ \left(
\begin{array}{c} -C\\
B\\ - A
\end{array} \right) \rightarrow \left(
\begin{array}{c} C\\
- B\\ A
\end{array} \right).$
Thus, we get a massless eigenstate $ |\  0\ \rangle = \fr 1 L
\left(
\begin{array}{c} C\\
- B\\ A
\end{array} \right).$
Now we look for other eigenstates
 \bea \left(
\begin{array}{ccc} 0 & A & B
 \\ - A & 0 & C \\ -B
 & -C  &0
\end{array} \right)
 \left(
\begin{array}{c} x\\
y\\ z
\end{array} \right) = \pm i L \left(
\begin{array}{c} x\\
y\\ z
\end{array} \right)
\label{lin18}  \eea or $ Ay  + B z =  (\pm i L) x $ and $
 - Ax  + C z = (\pm i L y)
$ or $ \pm i L x - Ay  = B z $ and $
  A x  \pm i L y = C z
$.  We get then, \be \left( \fr{ BC -   i AL}{AB+iCL},\hs
\fr{A^2+C^2}{AB+iCL}, \hs 1\right)\label{lin19}  \ee for $\la = i
L$, and
 \be \left( \fr{- BC -   i AL}{- AB+iCL},\hs -\fr{A^2+C^2}{-AB+iCL},
\hs 1\right)\label{lin19}  \ee for $\la = - i L$. Thus  the
normalized eigenstates are given by \bde
 \bea |\  \phi_+ \  \rangle^T & =&
\fr{1}{\sqrt{2(A^2+C^2)(A^2+B^2+C^2)}} (BC-iAL, A^2+C^2,
AB+iCL),\\
|\  \phi_- \  \rangle^T & =&
\fr{1}{\sqrt{2(A^2+C^2)(A^2+B^2+C^2)}} (BC+iAL, A^2+C^2, AB - iCL)
\label{lin20} \eea \ede

Now we have obtained three eigenstates $  |\  \phi_0 \ \rangle , \
|\  \phi_+ \  \rangle , \  |\  \phi_- \  \rangle $. It is
straightforward to check that they form an orthonormal basis. Note
that, for an anti-Hermitian matrix like $M$, its eigenstates (with
different $E$) are also orthogonal!

Now, it is straightforward to construct the eigenstates of the
true Hamiltonian $H$. For $E \neq 0$, Eq. (\ref{lin8}) gives \be
\vph_B = -\fr 1 E M \vph_A \label{lin21}  \ee \ben \item {\it For
$E=L$}: $ \vph_B = -\fr 1 L M \vph_A$. Taking $\vph_A = \phi_+$,
then $\vph_B =  -\fr i L  L \phi_+ = -i \phi_+$. Thus we have $
\Psi_1 = \fr{1}{\sqrt{2}}  \left(
\begin{array}{c} \phi_+ \\
-i \phi_+
\end{array} \right)$.
 Similarly, let  $\vph_A = \phi_-$,  then $\vph_B =  -\fr 1 L
(-i L) \phi_- = i \phi_-$. Hence $ \Psi_2 = \fr{1}{\sqrt{2}}
\left(
\begin{array}{c} \phi_- \\
i \phi_-
\end{array} \right)$.
Note that $ \langle \Psi_1 | \Psi_2 \rangle = \fr 1 2 (\phi^+_+, i
\phi^+_+)\left(
\begin{array}{c} \phi_- \\
i \phi_-
\end{array} \right) = \phi^+_+ \phi_- (1-1) = 0$.
 \item {\it For $E= - L$}:
$ \vph_B = \fr 1 L M \vph_A $. Analogously, we get
 $ \Psi_5 = \fr{1}{\sqrt{2}}
\left(
\begin{array}{c} \phi_+ \\
i \phi_+
\end{array} \right) \hs  {\rm and} \hs
\Psi_6 = \fr{1}{\sqrt{2}} \left(
\begin{array}{c} \phi_- \\
- i \phi_-
\end{array} \right).$
 \item {\it For $E= 0$}: a special case
$ \Psi_3 =  \left(
\begin{array}{c} \phi_0 \\
0
\end{array} \right)  \hs  {\rm and} \hs
\Psi_4 =  \left(
\begin{array}{c} 0 \\
\phi_0
\end{array} \right).$
\een
 Now we only need to rewrite the perturbation $V$ in a
new basis \bea V_{ij} =  \langle \Psi_i | V |  \Psi_j \rangle ,\hs
i,j = 1,2,...,6,  \hs V = \left(
\begin{array}{cc}
 0  &  0
 \\
0 & \De
\end{array} \right).
\label{lin26}  \eea
 We also have
$ \De  = \sum_{i,j} | \phi_i  \rangle \De_{ij}  \langle \phi_j |.$
 Sine $\Psi_i$ are expressed through just three functions
 $\phi_+,\  \phi_0,\  \phi_-$, so  we only need to work out the $ 3 \times 3$
 matrix!
 Finally, the $6 \times 6 $ matrix is
\bea \left(
\begin{array}{cccccc} \fr 1 2 \De_{++}  & -\fr 1 2 \De_{+-} & 0 &
\fr{i}{\sqrt{ 2}} \De_{+0}& - \fr 1 2 \De_{++}& \fr 1 2 \De_{+-}
 \\ -\fr 1 2 \De_{-+} &  \fr 1 2 \De_{--}  & 0 & -\fr{i}{\sqrt{ 2}}
 \De_{-0}& -\fr 1 2 \De_{-+} & -\fr 1 2 \De_{--}  \\
0&0&0&0&0&0\\ -\fr{i}{\sqrt{ 2}} \De_{0+}& \fr{i}{\sqrt{ 2}}
\De_{0-}&0 &\De_{00}&\fr{i}{\sqrt{ 2}} \De_{0+}&-\fr{i}{\sqrt{ 2}}
\De_{0-}\\ - \fr 1 2 \De_{++} &\fr 1 2 \De_{+-}&0 &-\fr{i}{\sqrt{
2}} \De_{+0}&\fr 1 2 \De_{++}&-\fr 1 2 \De_{+-}\\
 \fr 1 2 \De_{-+} &-\fr 1 2 \De_{--}&0 &\fr{i}{\sqrt{ 2}}
\De_{-0}&-\fr 1 2 \De_{-+}&\fr 1 2 \De_{--}
\end{array} \right)\nn
\eea To calculate the leading corrections, only
need to diagonalize the $2\times 2 $ matrix. Note that $\De_{ij} =
\De_{ij}(d,s,t)$ are homogeneous function of order 1 for
$(d,s,t)$. The matrix $ \left(
\begin{array}{cc}
 0  &  0
 \\
0 & \De_{00}
\end{array} \right)$,
 gives $ \De E = 0,\  \De_{00} + {\cal O} (g^2)$.
 which is analytic function. So, the energy spectra for  $E_3,
E_4$ are given \bea  E_3&=& 0 + {\cal O} (g^2),\  E_4  = \De_{00}
+ {\cal O} (g^2)
 \label{lin30}
 \eea
 Next, the matrix
$ \left(
\begin{array}{cc}
 \fr 1 2 \De_{++}   & - \fr 1 2 \De_{+-}
 \\
-\fr 1 2 \De_{-+}  & \fr 1 2 \De_{--}
\end{array} \right),$
 has two eigenvalues given by
 \be
 \De E  = \fr{\De_{++} +\De_{--}}{4} \pm
\sqrt{\left(\fr{\De_{++} -\De_{--}}{4}\right)^2
+\fr{|\De_{+-}|^2}{4}} \label{lin31}\ee which is, in general,
nonanalytic!. For instance, $f(d,s,t) = \sqrt{d^2+s^2+t^2}$ is
{\it not} analytic at $(d,s,t) = (0,0,0)$. Thus, we obtain \bde
 \bea
E_{1,2}& =& L + \fr{\De_{++} +\De_{--}}{4} \pm
\sqrt{\left(\fr{\De_{++} -\De_{--}}{4}\right)^2
+\fr{|\De_{+-}|^2}{4}} +  {\cal O} (g^2),\\
E_{5,6}& =&  - L + \fr{\De_{++} +\De_{--}}{4} \pm
\sqrt{\left(\fr{\De_{++} -\De_{--}}{4}\right)^2
+\fr{|\De_{+-}|^2}{4}} +  {\cal O} (g^2)
 \label{lin32}
 \eea
 \ede
 The $ {\cal O} (g)$ corrections are nonanalytic in general!.
 To get explicit results we  calculate $\De_{ij}$.\\
 $ \De_{++} = \phi_+^+ \De \phi_+ = \fr{1}{2(A^2+C^2)L^2}\\
  \times (
 BC + i AL, A^2+C^2, AB - i CL) \left(
\begin{array}{ccc} 0 & d & s
 \\ d & 0 & t \\ s
 & t  &0
\end{array} \right)
 \left(
\begin{array}{c} BC-iAL\\
A^2+C^2\\  AB+iCL
\end{array} \right) = \fr{1}{L^2} (B C d  - A C s + A B t).$
 Noting $
 \phi_- = \phi_+^*$,
  we have
 $\De_{--}= \De_{++}$.

 Now we turn our attention to $\De_{00}$.
$  \De_{00} = \phi_0^+ \De \phi_0 = \fr{1}{L^2} ( C, -B, A)
  \left(
\begin{array}{ccc} 0 & d & s
 \\ d & 0 & t \\ s
 & t  &0
\end{array} \right)
 \left(
\begin{array}{c} C\\
-B\\  A
\end{array} \right)
 = - \fr{2}{L^2} (B C d  - A C s + A B t) = - 2 \De_{++}$.\\
 $\De_{+-} = \phi_+^+ \De \phi_- =
 \fr{1}{2(A^2+C^2)L^2} \times (
 BC + i AL, A^2+C^2, AB - i CL) \left(
\begin{array}{ccc} 0 & d & s
 \\ d & 0 & t \\ s
 & t  &0
\end{array} \right)
 \left(
\begin{array}{c} BC+iAL\\
A^2+C^2\\  AB-iCL
\end{array} \right)
= \left(\fr{ BC + i AL}{L^2}\right) d + \left[\fr{AC( B^2+L^2) - i
BL(C^2 - A^2)}{(A^2+C^2)L^2}\right]s+  \left(\fr{AB - i
CL}{L^2}\right)t$.

We have $ \De_{-+} = (\De_{+-})^*$. Similarly
 $\De_{0+} = \phi_0^+ \De \phi_+ = \left[\fr{ C(L^2 - 2B^2) +
 i ABL}{\sqrt{2(A^2+C^2)}L^2}\right] d
+ \left[\fr{2ABC +i L(C^2-A^2)}{\sqrt{2(A^2+C^2)L^2}}\right]s +
\left[\fr{A(L^2-2B^2) - i LBC}{\sqrt{2(A^2+C^2)L^2}}\right]t,$
 and\\
 $\De_{0-} = \phi_0^+ \De \phi_- = \left[\fr{ C(L^2 - 2B^2) -
 i ABL}{\sqrt{2(A^2+C^2)}L^2}\right] d
+ \left[\fr{2ABC -i L(C^2-A^2)}{\sqrt{2(A^2+C^2)L^2}}\right]s
+\left[\fr{A(L^2-2B^2) + i LBC}{\sqrt{2(A^2+C^2)L^2}}\right]t$.

 Noting that $\phi_0$ is real, and $\phi_+ = \phi_-^*$, we have
 the following properties
 $ \De_{0-}=\De_{0+}^* =  \De_{-0}^* =\De_{+0}$.
 To complete, we calculate
\bde
 \bea
 |\De_{+-}|^2 & = & \fr{1}{L^4 (A^2 + C^2)^2}\left\{(A^2 + C^2)^2 [ (B^2
 C^2 + A^2 L^2) d^2  + (A^2
 B^2 + C^2 L^2) t^2\right.\crn
 &-& 2 (A^2 + C^2) A C dt] + (A^2 B^2 + C^2 L^2)(B^2
 C^2 + A^2 L^2) s^2\crn
 &+& \left.2 (A^2 + C^2) [ (B^2 C^2 + A^2 L^2) A B d s +
 (A^2 B^2 + C^2 L^2)B C s t]\right\}\label{38}
\eea \ede

\end{document}